\documentclass[12pt,a4paper,subeqn]{article}
\usepackage[tight,hang]{subfigure}

\usepackage{psfrag}
\usepackage{graphicx}
\usepackage{amsmath}     
\usepackage[square,comma,sort&compress]{natbib} 
\usepackage{geometry}
\usepackage{wrapfig}
\title {Computer simulation of two continuous spin models using
Wang-Landau-Transition-Matrix Monte Carlo Algorithm 
\\[8mm]}
\author{Shyamal Bhar
\footnote{E-mail address: sbhar@research.jdvu.ac.in }\,\,\,  and Soumen Kumar Roy
\footnote {Corresponding author. E-mail address: skroy@phys.jdvu.ac.in,
Telephone No: +91 9331910161,
Fax: +91 33 24146584}\\
Department of Physics, Jadavpur University,\\[1mm]
Kolkata - 700 032, INDIA}
\date{  }
\begin{document}

\maketitle
\begin{abstract}
\noindent Monte Carlo simulation using a combination of Wang Landau (WL) and
Transition Matrix (TM) Monte Carlo algorithms to simulate two lattice
spin models with continuous energy is described. One of the models, the
one-dimensional Lebwohl-Lasher model has an exact solution and we have
used this to test the performance of the mixed algorithm (WLTM). The other
system we have worked on is the two dimensional XY-model. The purpose 
of the present work is to test the performance of the WLTM algorithm in 
continuous models and to suggest methods for obtaining best results
in such systems using this algorithm.
\end{abstract}


%%%%%%%%%%%%%%%%%%%%%%%%%%%%%%%%%%%%%%%%%%%%%%%%%%%%%%%%%%%%

\textheight=8.9in
\section {\bf Introduction}
\noindent The Metropolis algorithm \citep{metro}, proposed 
more than half a century
ago has proved to be a very important and useful tool in Monte Carlo (MC)
simulation, for the prediction of equilibrium thermodynamic properties
of a large variety of systems. In this method one generates a large
number of microscopic states of a system according to their canonical
probabilities and the averages of different observables are calculated
directly. However, problems crop up in the neighbourhood of a critical
point, where due to critical slowing down \cite{newman} it 
becomes rather difficult
to overcome the large correlation time problem and get reliable results.
Also for simulating systems with difficult potential energy landscapes
as in proteins, spin-glasses etc, simulation using Metropolis algorithm
results in trapping of the random walker in a potential energy minimum
resulting in rather inefficient sampling.\\
\noindent During the last couple of decades or so a number of 
MC algorithms \cite{wlprl, yama,herrmann, rathore, jain, jayasri, 
pablo, errin, swendsen, shell, ghulgha} have
been proposed which are designed to overcome the shortcomings of the 
Metropolis algorithm. To deal with the critical slowing down in a discrete
spin model a cluster algorithm was proposed by Swendsen and Wang \cite{s_wang},
while the Wolff cluster algorithm \cite{wolff} has proved to be an 
important tool
for simulating continuous spin models. The multiple histogram
reweighting technique of Ferrenberg and Swendsen \cite{ferren2} 
emerged as a very
useful method of improving the results obtainable from  MC simulations 
in general.\\
\noindent Instead of tracing out the historical development of the MC
algorithms, detailed discussion of most of which may be found in recent
books like Newman and Berkema \cite{newman} and Landau and Binder 
\cite{binder}, we mention two algorithms which are of particular
relevance to the subject of this communication. One of these is the
Transition Matrix Monte Carlo (TMMC) algorithm developed by Oliveria
et al \cite{oliveria1} and subsequently generalized by 
Wang and co-workers \cite{swendsen}. The other algorithm, which was
developed by Wang and Landau \cite{wlprl} and goes by their
name (WL), has drawn wide attention of researchers since its 
inception \cite{shell2, yan, landau, xu}. Both of these 
methods determine the density of states
(DOS) of a system as a function of one or more macroscopic
observables like energy, order parameter, correlation function etc.
A knowledge of the DOS enables one to determine the partition function
or the averages of other thermodynamic quantities by the usual Boltzmann
reweighting procedure. Both TMMC and WL algorithms depend on broad sampling
of the phase space and are easy to implement. While the WL algorithm
is capable of an efficient sampling of the phase space, the TMMC
method gives more accurate estimates of the DOS and this improves with
longer MC runs.\\
\noindent Shell and co-workers in 2003 \cite{shell} proposed an algorithm
which combines the TMMC and WL algorithms in an efficient way so as to make
best use of the benefits of each algorithm. The new algorithm, known
as the Wang Landau Transition Matrix (WLTM) Monte Carlo algorithm, uses
the WL sampling scheme along with the TMMC method so as to result in an
algorithm which is both efficient and accurate. The method was tested by
Shell et. al \cite{shell} in two dimensional Ising model and a Lenard-Jones
fluid which suggests that the method is capable of handling both discrete
and continuous systems.\\
\noindent In this communication we have performed elaborate WL 
and WLTM simulation
of two continuous spin models. Continuous models are understandably more
laborious and difficult subjects of MC simulation and need special
considerations regarding the choice of certain parameters --- a problem
one does not encounter while simulating a discrete model. 
The aim of this paper is to critically judge the performance
of WL and WLTM methods in continuous systems. We have investigated
the manner in which the performance of the WLTM algorithm in particular 
could be
improved in such systems. We have chosen $i)$ the  one dimensional
Lebwohl-Lasher (1-d LL) model \cite{lebwohl} and $ii)$ the two 
dimensional XY (2-d XY) model
for this purpose. The reason for the choice of the former is that
the model is exactly solvable \cite{romerio} and 
hence one can compare the
performance of the WLTM algorithm with the exact results. This is analogous 
to the practice where the results of a new MC algorithm is being compared
with the exact results on a 2d Ising model. For the XY-model, where
the exact partition function is not available, we have compared our
results with the results of the conventional Metropolis simulation.\\
\noindent The scheme of the presentation is as follows. The WL, TMMC
and WLTM algorithms are briefly described in the next section. In section
3 we have described the models we have worked upon. Computational details
appear next and is followed by the results we have obtained.
\section {\bf The different algorithms}
\subsection {\bf The Wang Landau algorithm}
\noindent In a system where the energy can take up continuous values, 
discretization of the energy range is necessary to label the macrostates 
of a system. The allowed energy range is divided into a number of bins, and
the mean energy of the $I^{th}$ bin is taken as $E_I$. The WL algorithm
directly determines the density of states (DOS) of the system as a function
of $E_I$ which is precisely the number of microscopic configurations which
have energy lying in the range which corresponds to the $I^{th}$ bin. This 
therefore is just the degeneracy factor which is denoted by $\Omega(I)$
 and it is natural to consider bin-widths which are sufficiently small.
Of course one can perform WL simulation to determine the DOS as a function of 
quantities other than energy like, say, the order parameter, correlation
function etc. Also, a determination of DOS (more precisely, joint density
of states, JDOS) as a function of two variables is possible and is necessary
in many situations \cite{troster, kisor, zhou1, poulain}.\\
\noindent For a macroscopic system, $\Omega (I)$ is a large number, and it
is convenient to work with its logarithm, $g(I)=ln \Omega (I)$. At the
beginning of the simulation one has no knowledge of the $g(I)$'s --- these 
are therefore set equal to zero for all values of $I$ and the algorithm 
generates 
the DOS profile, which progressively becomes closer to the actual DOS of the
system, by an iterative process, which is briefly outlined below. Since
the DOS is independent of temperature, and contains complete 
information about the system, the task is to determine it (or the JDOS, if 
necessary) as accurately as possible. The rest of the work, which involves
the determination of the partition function $Z$ or other thermodynamic 
quantities at any temperature is done in a simple and straight forward
manner by the standard Boltzmann reweighting procedure.\\
\noindent The implementation of the WL algorithm is done in the following
manner. One starts with some microscopic configuration and a random walk
is generated by rotating the spins (one at a time, in our case). Thus
starting from a microstate $i$ one generates another state
$j$. The probability with which the transition $i\rightarrow j$
is accepted is given by,
\begin{equation} \label{pe1e2}
P(i, j) = min \left(\frac{\Omega(I)}{\Omega(J)},1 \right)
\end{equation}
\noindent where $i\in I$ and $j\in J$ relates the microstates with the 
corresponding macrosates.
Thus the probability of acceptance is inversely proportional
to the current value of the DOS. In the event of the new state being 
accepted one makes the following modifications:
\begin{subequations} \label{accep}
\begin{equation}
g(J) \rightarrow  g(J)+lnf
\end{equation}
\begin{equation}
\text{and}\qquad H(J)  \rightarrow  H(J)+1 
\end{equation}
\end{subequations}
\noindent Here $H(I)$ represents the histogram count of the $I^{th}$ bin
and, $f$ is a modification factor whose initial value is greater than 1 (
in our case we started with $lnf=1$). In case the probability test 
(\ref{pe1e2}) fails
, the changes given in equation (\ref{accep}) are made to the bin labelled 
by $I$ 
instead of J. This procedure
is continued till the histogram becomes sufficiently flat, which may be say 
$90\%$. This means that the histogram count $H(I)$ for each value of $I$
is at least $90\%$ of the average $\left(\sum^{M}_{I=1}H(I)\right)/M$, where
$M$ is the number of bins. When this condition is fulfilled, one iteration
is said to be complete. One then resets the $H(I)$'s
to zero for each value of $I$, changes 
$lnf\rightarrow lnf/2$ (or in some other way) and starts a fresh 
iteration using 
the $g(I)$'s which have been generated in the previous iteration. Iterations
are continued with the DOS steadily approaching the true DOS profile, till
$lnf$ is sufficiently small. We have stopped the process when $lnf$ is as 
small as $10^{-9}$. The error which is present in the density of states
has been predicted to be proportional to $\sqrt{lnf}$ as is apparent from the
theoretical work of Zhou and Bhatt \cite{zhou} . This has been tested for 
a number
of discrete and continuous models and the 
prediction has been found to be correct \cite{lee, suman}.
\subsection{\bf The Transition Matrix Monte Carlo algorithm}
\noindent The TMMC algorithm also directly evaluates the DOS of a system
and was first proposed by Oliveria {\it et al.} \cite{oliveria1} 
in the year 1996. Let $I$ and $J$ be the
labels of two macrostates and $i$ and $j$
represent the set of all microstates which correspond to the states $I$ and
$J$ respectively. The probabilities for the transitions  
$i \rightarrow j$
and $I \rightarrow J$ are written as $t(i,j)$ and $T(I,J)$ respectively.
Both these quantities must satisfy the following conditions:
\begin{subequations} \label{tmnorm}
\begin{equation}
\sum_j{t(i,j)}=1, \qquad t(i,j)\ge 0 
\end{equation}
\begin{equation}
\text{and}\qquad  \sum_J{T(I,J)}=1, \qquad T(I,J)\ge 0
\end{equation}
\end{subequations}
\noindent The two types of transition probabilities are related in a simple 
manner. If $\Omega(I)$ and $\Omega(J)$ are the DOS's corresponding to the 
macrostates $I$ and $J$, then one can write,
\begin{equation}\label{tprob}
T(I,J)=\frac{1}{\Omega(I)}\sum_{i\in I}\sum_{j\in J}{t(i,j)}
\end{equation}
\noindent If $T(J,I)$ is the reverse transition probability \textit{i.e.}
for the $J\rightarrow I$ transition, then from eqn (4) it follows that,
\begin{equation}\label{tmratio1}
\frac{T(I,J)}{T(J,I)}=\frac{\Omega(J)}{\Omega(I)}
\frac{\displaystyle{\sum_{i\in I}\sum_{j\in J}{t(i,j)}}}{\displaystyle
{\sum_{j\in J}\sum_{i\in I}{t(j,i)}}}
\end{equation}
\noindent The transitions like $i\rightarrow j$ etc. are effected in a MC
simulation by random walk in the configuration space and in lattice-spin
models, and are constituted by spin flips (in discrete model) or spin
rotations (in continuous models). The process of transition $i\rightarrow j$
is actually composed of two parts: $i)$ the transition being proposed and
$ii)$ it being accepted. If we represent by $a(i,j)$, the probability of
proposing the transition $i\rightarrow j$ and by $P(i,j)$, the
probability of this being accepted then we can write,
\begin{equation}\label{tp}
t(i,j)=a(i,j)P(i,j)
\end{equation}
\noindent The probability $a(i,j)$ depends on the type of MC move used
while the quantity $P(i,j)$ has the flexibility of being chosen. If one 
considers an infinite temperature then $P(i,j)=1$ for all states
$i$ and $j$. This can easily be seen to be correct by considering a 
Metropolis dynamics at infinite temperature (then $P(i,j)$ would simply be
the ratio of the Boltzmann factors of the two states ). Therefore combining
equations (\ref{tmratio1}) and (\ref{tp}) for the temperature $T=\infty$, 
we can write,
\begin{equation}\label{tmratio2}
\frac{T_{\infty}(I,J)}{T_{\infty}(J,I)}=\frac{\Omega(J)}{\Omega(I)}
\frac{\displaystyle{\sum_{i\in I}\sum_{j\in J}a(i,j)}}
{\displaystyle{\sum_{j\in J}\sum_{i\in I}a(j,i)}}
\end{equation}
\noindent where $T_{\infty}(I,J)$ represents the infinite temperature
transition probability. It may be noted that the DOS's which appear in the 
r.h.s. of equation (\ref{tmratio2}) remain unchanged for these are not 
dependent on 
temperature. For random walk in the configuration space involving symmetric
moves, like single spin flip or rotation, as we have done, $a(i,j)=a(j,i)$
and equation (\ref{tmratio2}) can be simplified to,
\begin{equation}\label{tw}
\frac{T_{\infty}(I,J)}{T_{\infty}(J,I)}=\frac{\Omega(J)}{\Omega(I)}
\end{equation}
\noindent This equation indicates that one can calculate the DOS from a 
knowledge of the infinite temperature transition probabilities. The 
procedure for this is to keep a record of a transition matrix, called the
C-matrix, $C(I,J)$ for all proposals $I\rightarrow J$, during the 
random walk. One starts the simulation with $C(I,J)=0$ for all $I,J$ and 
updates the matrix by 
\begin{equation}\label{cupdate}
C(I,J) \to C(I,J)+1
\end{equation}
\noindent for each proposal $I\rightarrow J$. Once the construction of the
C-matrix is started it is never zeroed during the simulation. So the current
estimate of the infinite temperature transition probability 
($\widetilde{T}_{\infty}$) is given by,
\begin{equation}\label{ttilde}
\widetilde{T}_{\infty}(I,J)=\frac{C(I,J)}{\displaystyle{\sum_{K}C(I,K)}}
\end{equation}
\noindent where the sum extends over all $K$ for which the $I\rightarrow K$
transition can occur. Of course, the C-matrix element, connecting
states between which a transition can never be proposed is always zero.
Finally with the knowledge of 
$\widetilde{T}_{\infty}(I,J)$ one can estimate the DOS ($\Omega$) from
equation (\ref{tw}) and this is possible at any stage of the simulation. 
However,
equation (\ref{tw}) leads to an over specified condition for the purpose of 
determining the $\Omega$'s as, for a given initial state $I$, a number of 
possible states $J$ can be proposed and each such pair $(I,J)$ 
satisfies equation
(\ref{tw}) leading to a number of equations for determining $\Omega(I)$, say.
One can then, for the best estimate of $\Omega(I)$, make a multi-variable
optimization as has been described in \cite{swendsen, shell}. For 
most simulations however,
it is adequate only to consider transitions involving $I,J$ which are
neighbouring states. 
In the TMMC algorithm, 
in order to ensure that the
random walker visits all macrostates in the region of interest, a uniform
ensemble, where all macrostates are equally probable is considered. The 
probability of occurrence of a given microstate $i$ is then inversely 
proportional to the number of microstates which are associated with the
macrostate $I$, where $i\in I$. So the probability of acceptance of a move
$i\rightarrow j$ is given by,
\begin{equation}\label{pacc}
P(i,j)=min\left(\frac{\Omega(I)}{\Omega(J)},1\right)
\end{equation}
\noindent Since the density of states are not known a priori, the above 
acceptance criterion is implemented by using equation (\ref{tw}) for 
the ratio of 
the density of states:
\begin{equation}\label{pacctilde}
P(i,j)=min\left(\frac{\widetilde{T}_{\infty}(J,I)}
{\widetilde{T}_{\infty}(I,J)},1\right)
\end{equation}
\noindent It may be noted that in the above mentioned procedure one does not
actually use the strict acceptance criterion for all proposed moves. This 
does not lead to any self contradiction in what has been stated in the
arguments leading to equation (\ref{pacctilde}). This is because 
the manner in which
one constructs the C-matrix, from which the estimate of the infinite-T 
transition probability is generated via equation (\ref{ttilde}), 
is independent of the actual
acceptance probability used and depends only on the configurations 
which are being
proposed. The errors in the TMMC simulation for different models have
been described in detailed in the work of Wang and Swendsen \cite{swendsen}.
\noindent However, the TMMC algorithm suffers from a number of drawbacks. The
most important of these is that the convergence of the density of states
in this method is not guaranteed and consequently the method is inefficient
in terms of the computer time necessary.
\subsection{\bf The Wang Landau-Transition Matrix Monte Carlo algorithm}
\noindent In the WLTM algorithm, first proposed by Shell and co-workers
\cite{shell} one efficiently combines the WL and TM Monte Carlo methods. 
Here one
uses the original acceptance criterion given by equation (\ref{pacc}) and the 
density of states and a histogram count are updated via equation 
(\ref{accep}).
Along
with this, a C-matrix, which is never zeroed, is constructed and this may 
start at a suitable stage of the simulation process (as is elaborated
in section (5) below ). The DOS available from the C-matrix
via the infinite-T transition probability ( equations (\ref{ttilde}) 
and (\ref{tw}) ) is used to 
replace the DOS which is generated by the WL-algorithm. This we call the
'refreshing' of the DOS and is done after each iteration. The result of 
amalgamating the WL and TM algorithms is to yield a  two fold advantage:
$i)$ better sampling of states due to the WL algorithm and $ii)$ improvement
in the accuracy of DOS as a result of the TMMC algorithm.
\section{\bf The models used in the present work}
\subsection{\bf The 1-d Lebwohl-Lasher model}
\noindent In this model one considers a linear-array of three dimensional
spins interacting with nearest neighbours (nn) via a potential,
\begin{equation}\label{vll}
V_{ij}=-P_2(cos\theta_{ij})
\end{equation}
\noindent where $P_2$ is the second Legendre Polynomial and $\theta_{ij}$
is the angle between the nearest neighbour spins $i,j$. Periodic boundary
conditions are used. This model represents an one-dimensional nematic
liquid crystal. The 3-d Lebwohl-Lasher (LL) model is a lattice 
version of a 3-d nematic,
described in the mean field approximation by the Mair-Saupe 
theory \cite{maier1}, and exhibits
an orientational order-disorder transition. The 1-d LL model has been
simulated by \cite{zannoni} and has also been solved exactly 
by Vuillermot and Romerio \cite{romerio} 
using a 
group 
theoretic method. As is to be expected in a low-dimensional model with nn
interaction, the 1-d LL model does not exhibit any finite temperature phase
transition. The results obtained in \cite{romerio} are quoted below.
\noindent The partition function $Z_N(\widetilde{K})$ for the $N$ particle
system is given by,
\begin{equation}\label{pll}
Z_N(\widetilde{K})=\widetilde{K}^{\frac{N}{2}}exp[\frac{2}{3}N\widetilde{K}]
D^N(\widetilde{K}^{\frac{1}{2}})
\end{equation}
\noindent where $\widetilde{K}=\frac{3}{2T}$ is a dimensionless quantity. $D$
is the Dawson function given by,
\begin{equation}\label{dll}
D(x)=exp(-x^2)\int_{0}^{x}e^{u^2}\,du
\end{equation}
\noindent The dimensionless internal energy $U_N(\widetilde{K})$, entropy
$S_N(\widetilde{K})$ and the specific heat $C_N(\widetilde{K})$ are given by,
\begin{equation}\label{iell}
\frac{2U_N(\widetilde{K})}{N}=1+\frac{3\widetilde{K}^{-1}}{2}-\frac{3}{2}
\widetilde{K}^{-\frac{1}{2}}D^{-1}(\widetilde{K}^\frac{1}{2}).
\end{equation}

\begin{equation}\label{enll}
\frac{S_N(\widetilde{K})}{N}=\frac{1}{2}+\widetilde{K}-\frac{1}{2}\widetilde{K}
^\frac{1}{2}D^{-1}(\widetilde{K}^\frac{1}{2})+ln[\widetilde{K}^{-\frac{1}{2}}
D(\widetilde{K}^\frac{1}{2})].
\end{equation}

\begin{equation}\label{sll}
\frac{2C_N(\widetilde{K})}{N}=1-\widetilde{K}^\frac{3}{2}[\frac{\widetilde{K}^
{-1}}{2}-1]D^{-1}(\widetilde{K}^\frac{1}{2})-\frac{1}{2}\widetilde{K}
D^{-2}(\widetilde{K}^\frac{1}{2}).
\end{equation}
\subsection{\bf The 2-d XY model}
\noindent In this model planar spins placed at the sites of a planar
square lattice interact with nearest neighbours via a potential,
\begin{equation}\label{vXY}
V(\theta_{ij})=2\left\{1-\left[\cos^2(\theta_{ij}/2)\right]\right\}
\end{equation}
\noindent where $\theta_{ij}$ is the angle between the nearest neighbours
$i, j$. [This particular form of the interaction, rather than the more
conventional $-\cos(\theta_{ij})$ form, was chosen by Domany {\it{et. al}}
\cite{domany}
to enable them to modify the shape of the potential easily, which
led to what is now known as the modified XY-model ]. The XY-model is 
known to exhibit a quasi-long-range-order disorder
transition which is mediated
by unbinding of topological defects as has been described in the seminal
work of Kosterlitz and Thoules \cite{KT, K}. The XY-model has also been 
the subject
of extensive MC simulation over last few decades and some of the 
recent results may be found in \cite{palma}.
\section {\bf Computational details:}
\noindent In the 1-d LL model 
the configurations of the system are stored in terms of
the direction cosines of the spins, i.e. a set of three numbers 
($l_1, l_2, l_3$)
describe the orientation of each spin. The simulation is started with a random
configuration of the spins. To generate a new configuration, one spin at a
time is
randomly selected and a move $l_i \rightarrow l_i+p*r_i$ 
(for i=1, 2, 3)
 is effected where $r_i$ is a uniformly generated random number between -1 
and +1. To preserve the unit magnitude of the spins, the normalization 
condition $l^2_1+l^2_2+l^2_3=1$ is always applied.
In the 2-d XY-model,
two direction cosines are needed to describe
the orientation of each spin.
In the 1-d LL model, for a chain of length L (in units
of the lattice spacing), the energy of the system can have any value between
-L to L/2 while for the 2-d XY-model of linear dimension L the system
energy lies between 0 and $4L^{2}$.
In order to apply the WL or WLTM algorithm we have restricted the 
random walk in the energy space from -L to 0 and from 0 to $2L^{2}$ for the
1-d and 2-d models respectively 
(actually a small energy band
near the ground state was also excluded as is explained below), and this
energy range was divided into a number of bins, each having a width $d_e$. 
At the 
beginning of the simulation, we set all $g(I)$'s to zero. Starting from a 
microstate $i$, the random generation of configurations by the above
mentioned procedure leads to a new state $j$ being proposed. The probability 
of 
acceptance of this state is given by $P(i, j)$ in equation (1).
It may be noted that, 
there is always a possibility that
$j$ and $i$ represent the microstates belonging to the same macrostate --- in 
that 
case the move is always accepted.
The random walk is continued until the histogram
reaches 90\% flatness.
With the
knowledge of the C-matrix which has accumulated in the
process, we calculate the current DOS and use it to refresh the previous DOS.
This completes one iteration, following which, we reduce the modification 
factor $lnf\rightarrow lnf/2$, reset the histogram counts 
to zero and start a new 
random walk cycle. It is important to note that the C-matrix is never reset to
zero and hence it stores the entire 
history of the moves proposed in the simulation process.\\
\noindent The iterations are continued till the factor $lnf$ gets reduced to
$10^{-9}$. The WLTM algorithm has the flexibility in that, it is possible to
refresh the DOS from the accumulated C-matrix at any stage. It is even
possible to refresh at the very end when all the iterations have
been completed. This therefore amounts to doing pure WL simulation along with 
keeping a record of the C-matrix which is used only at the end to refresh the 
DOS. The dependence of the accuracy of the results resulting from refreshing 
the DOS at different stages is described in the following section.

\section{\bf Results and discussion}
\subsection{The 1-d LL model:}

\noindent We have simulated the 1-d model for system sizes $L= 80, 160$
and $220$
and the minimum of the energy range over which random walk has been carried
out were -79, -158, -218 respectively, the upper limit being 0 in each case.
The small cut in the energy near the ground state had to be made since it takes
a prohibitively long time to fill these states. Moves which take the system
energy outside the specified range are rejected and the C-matrix is updated
as $C(I,I) \rightarrow C(I,I)+1$ where the index $I$ labels the initial state.
%%%%%%%%%%%%%%%%%%%%%%%%%%%%%%%%%%%%%%%%%%%%%%%%%%%%%%
\begin{figure}[htbp]
\begin{center}
\resizebox{120mm}{!}{\rotatebox{0}{\includegraphics[scale=.45]{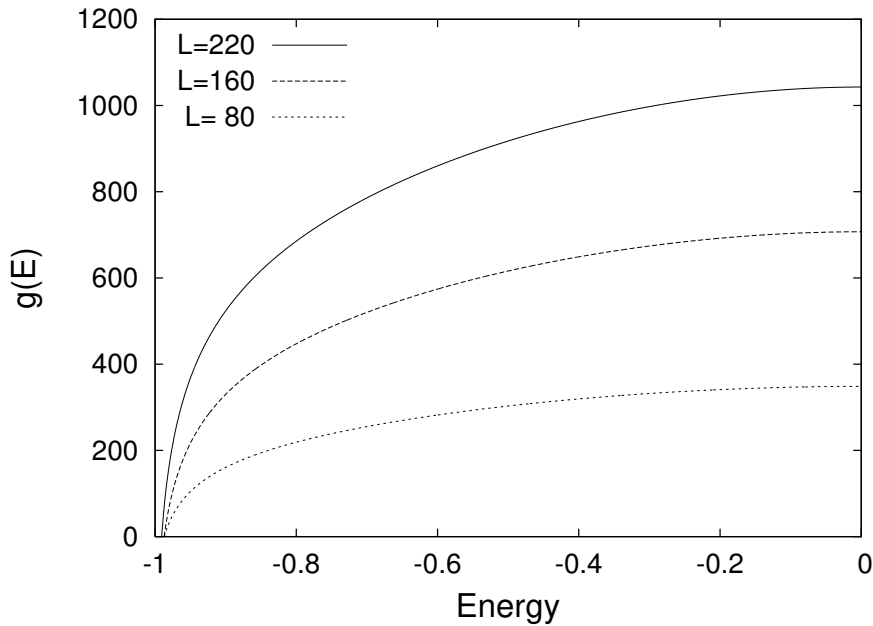}}}
\end{center}
\caption{ $g(E)$, the logarithm of the density of states
in the 1-d LL model
for the three system sizes are plotted against energy per particle
in the energy range of interest.}
\label{fig1.eps}
\end{figure}
%%%%%%%%%%%%%%%%%%%%%%%%%%%%%%%%%%%%%%%%%%%%%%%%%%%%%%%%
The energy
bin-width was chosen to be $d_e=0.1$ in all cases and the parameter $p$, defined
in the previous section, was also chosen to be 0.1. As the bin-width is made
narrower, the number of microstates which correspond to a particular bin gets
reduced.
Ideally, $d_e$ should be as small as possible for better 
discretization, but this is limited by computational difficulties in its 
implementation. Choosing a very small bin-width increases the number of bins
proportionately and the book-keeping work becomes difficult. Problems arise
both with CPU time as well as with available computer memory and 
this becomes severe 
particularly for a large lattice size. Optimization of these considerations
resulted in our choice of the bin width.\\
%%%%%%%%%%%%%%%%%%%%%%%%%%%%%%%%%%%%%%%%%%%%%%%%%%%%%%
\begin{figure}[htbp]
\begin{center}
\resizebox{120mm}{!}{\rotatebox{0}{\includegraphics[scale=.45]{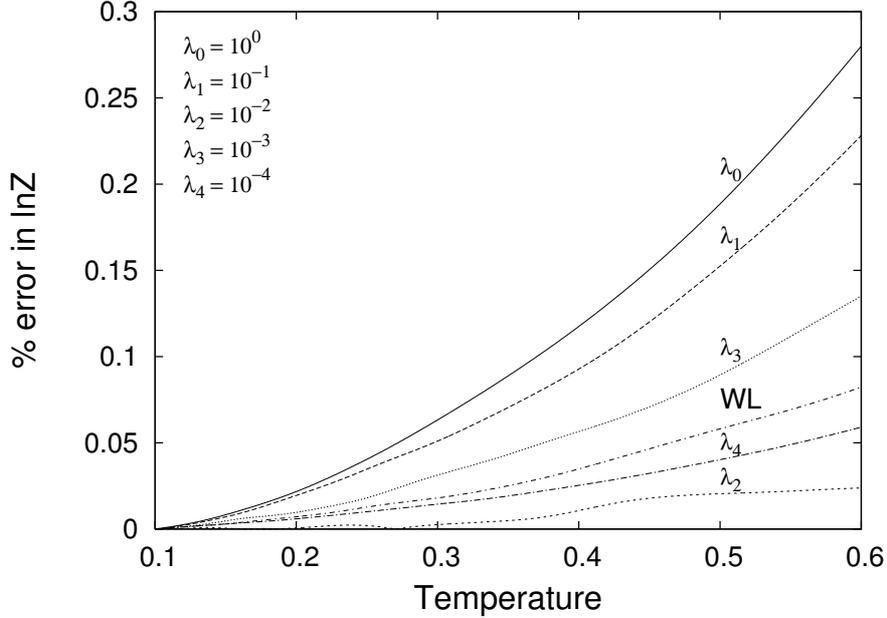}}}
\end{center}
\caption{  Percentage error in the logarithm of partition
function ($lnZ$) is plotted against temperature for the 1-d LL
model. Errors are plotted for the different cases where the
construction of C-matrix is started at different stages of simulation.
The meaning of the symbol $\lambda$ is described in the text.
Percentage error in $lnZ$
obtained from the WL algorithm with $90\%$ flatness criterion is also
shown in the figure.}
\label{fig2.eps}
\end{figure}
%%%%%%%%%%%%%%%%%%%%%%%%%%%%%%%%%%%%%%%%%%%%%%%%%%%%%%%%
\noindent The parameter $p$ determines the amplitude of a rotation which
is given to a spin at the time of generating new configurations. Too small
a value of $p$ leads to the configuration being trapped in the same bin
and a large value results in visits always outside the starting bin. Actually,
 while simulating a continuous spin system using WL or WLTM algorithm, a proper
 choice of the parameters $d_e$ and $p$ is particularly important. Clearly the
two parameters are interrelated and our choice of $p=0.1$ for the particular
value of $d_e=0.1$ was based on the consideration that about 50\% of the 
attempted moves go outside the starting bin so that care is taken of 
microstates in the initial bin as well as its neighbours \cite{shell}. 
This approach, 
besides
 ensuring a uniform sampling, also optimizes the CPU time.\\

%%%%%%%%%%%%%%%%%%%%%%%%%%%%%%%%%%%%%%%%%%%%%%%%%%%%%%
\begin{figure}[htbp]
\begin{center}
\resizebox{120mm}{!}{\rotatebox{0}{\includegraphics[scale=.45]{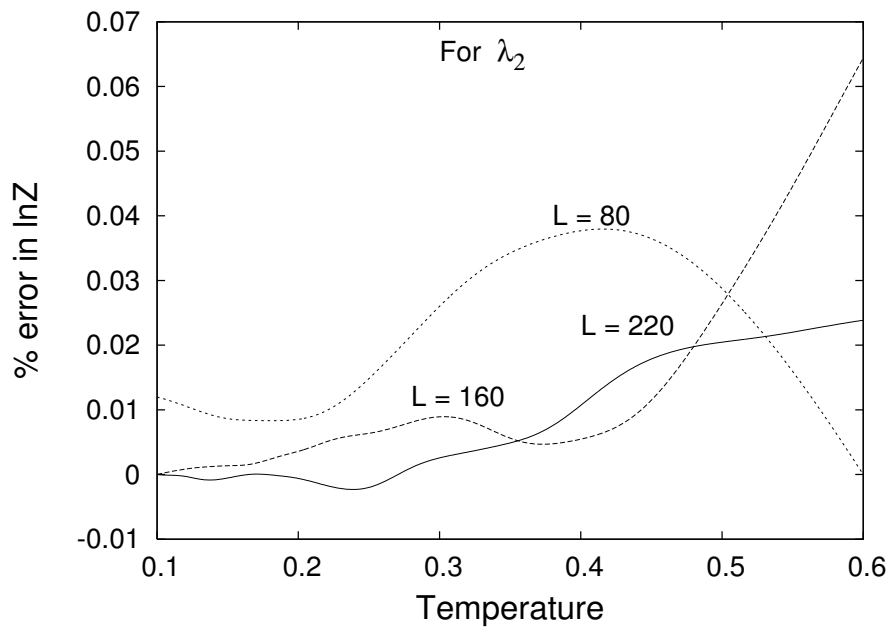}}}
\end{center}
\caption{ Percentage error in $lnZ$ is plotted against
temperature for three lattice sizes 80, 160 and 220.
The results are for the parameter $\lambda=\lambda_2$.}
\label{fig3.eps}
\end{figure}
%%%%%%%%%%%%%%%%%%%%%%%%%%%%%%%%%%%%%%%%%%%%%%%%%%%%%%%%
\noindent We have used the histogram flatness criterion to determine when an
iteration with a given modification factor $f$ is to be terminated. This 
condition, although it increases the CPU time enormously, was found to be 
necessary for minimizing the errors in the observables. Shell et al 
\cite{shell} while
prescribing the WLTM algorithm suggested that a single visit to each 
macrostate should be enough to determine where to terminate an iteration. We
have observed that while this works reasonably well for a model with a discrete
energy spectrum \cite{akash}, in a system with continuous energy spectrum, 
as is ours, a 
single visit criterion fails to work. This is however not surprising in view
of the fact that when discretization is achieved by a certain choice of the 
energy bin-width, the multiplicity of states associated with each bin (for
any choice of bin-width) is generally much greater than that for an
energy level in a discrete system of comparable size.\\

%%%%%%%%%%%%%%%%%%%%%%%%%%%%%%%%%%%%%%%%%%%%%%%%%%%%%%
\begin{figure}[htbp]
\begin{center}
\resizebox{120mm}{!}{\rotatebox{0}{\includegraphics[scale=.45]{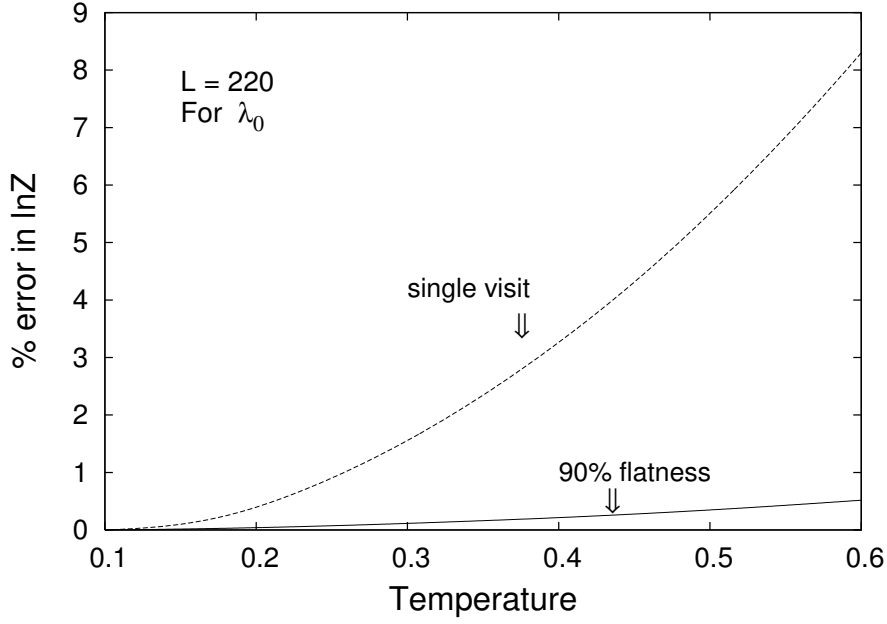}}}
\end{center}
\caption{ Percentage error in $lnZ$ is plotted against
temperature for three lattice sizes 80, 160 and 220.
The results are for the parameter $\lambda=\lambda_2$.}
\label{fig4.eps}
\end{figure}
%%%%%%%%%%%%%%%%%%%%%%%%%%%%%%%%%%%%%%%%%%%%%%%%%%%%%%
%%%%%%%%%%%%%%%%%%%%%%%%%%%%%%%%%%%%%%%%%%%%%%%%%%%%%%
\begin{figure}[htbp]
\begin{center}
\resizebox{120mm}{!}{\rotatebox{0}{\includegraphics[scale=.45]{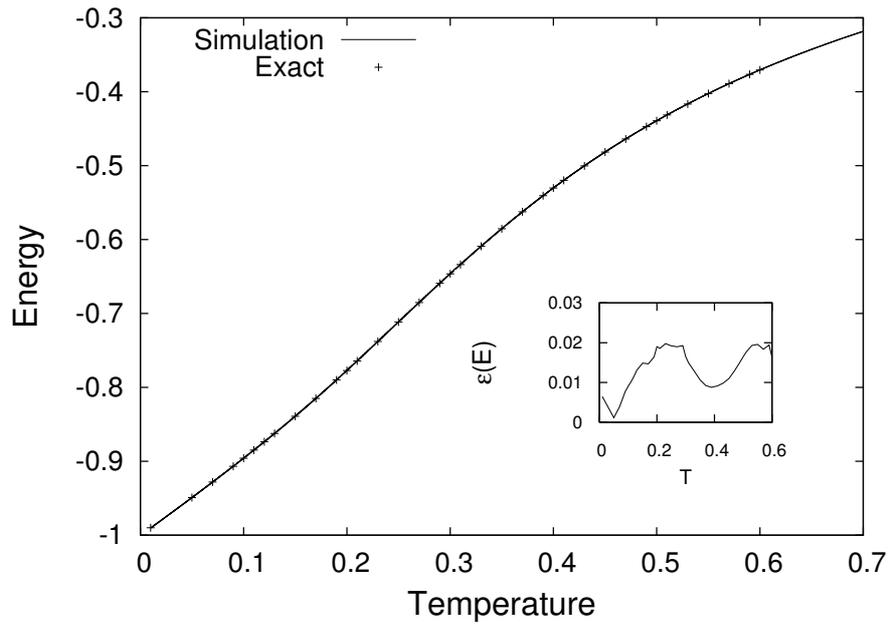}}}
\end{center}
\caption{The variation of energy per
particle in the 1-d LL model is plotted against temperature
(solid line) for $L=220$ and $\lambda_0$.
Exact results available for the system are also plotted in the same graph
and are indicated by '+' symbol. In the inset, the percentage error
$\epsilon (E)$ in energy is compared with exact results.}
\label{fig5.eps}
\end{figure}
%%%%%%%%%%%%%%%%%%%%%%%%%%%%%%%%%%%%%%%%%%%%%%%%%%%%%%%%
%%%%%%%%%%%%%%%%%%%%%%%%%%%%%%%%%%%%%%%%%%%%%%%%%%%%%%
\begin{figure}[htbp]
\begin{center}
\resizebox{120mm}{!}{\rotatebox{0}{\includegraphics[scale=.45]{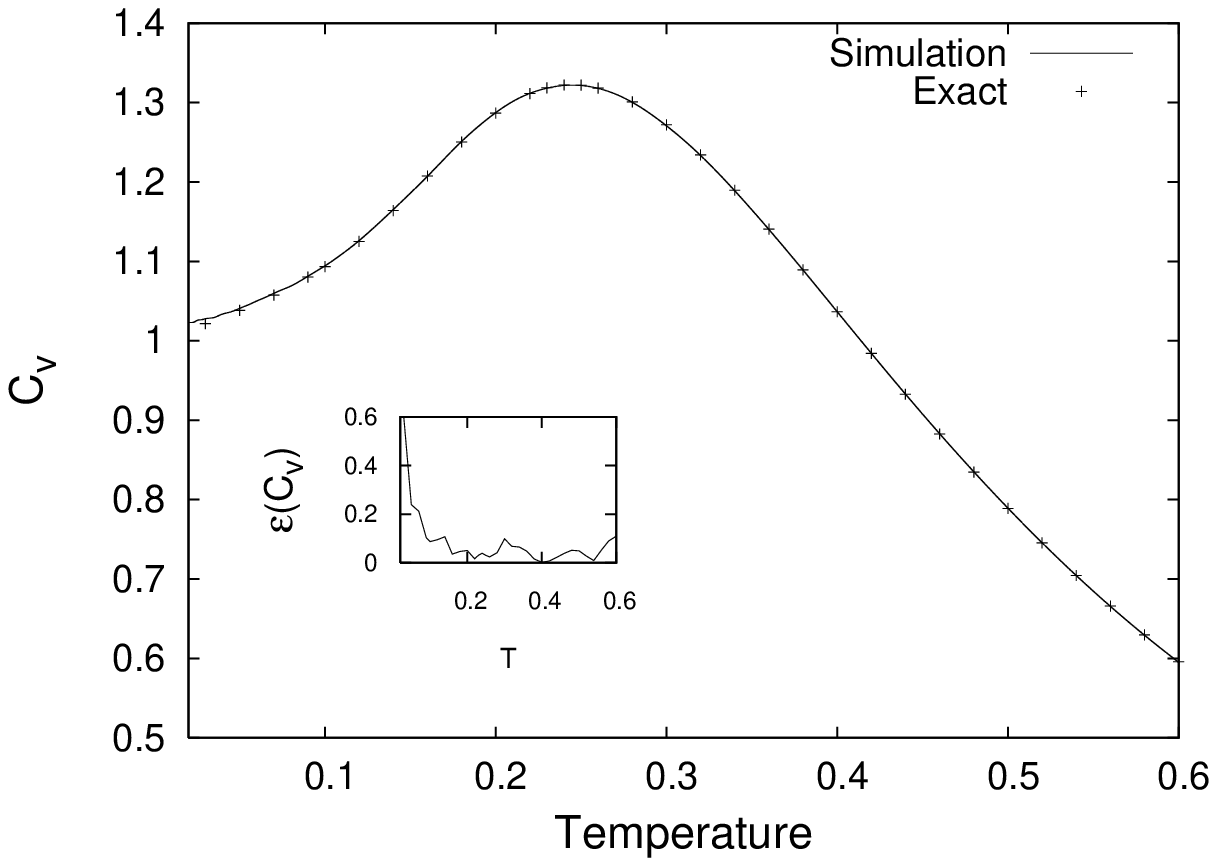}}}
\end{center}
\caption{ Specific heat per particle for the $L=220$
1-d LL model is plotted against
temperature for $\lambda=\lambda_0$. The '+' symbol indicates the exact
results for the system. The percentage error in the $C_v$ in comparison
with the exact results is shown in the inset.}
\label{fig6.eps}
\end{figure}
%%%%%%%%%%%%%%%%%%%%%%%%%%%%%%%%%%%%%%%%%%%%%%%%%%%%%%%%
\noindent In figure 1 we have shown the plots of $g(E)$, the logarithm
of the DOS against the energy per spin for three system sizes. Figure 2 is a 
plot of the error in lnZ, (Z being the partition function) and a comparison
has been made with the exact results of \cite{romerio}. 
Here we have presented the 
results of simulation where the construction of the C-matrix was started
at different stages of iteration. We denote by $\lambda_c$ a 
preassigned number and when the current value of $lnf$ becomes just less
than $\lambda_c$, the construction of the C-matrix is started.
For instance, the graph for 
$\lambda_c=\lambda_2=10^{-2}$ shows the error in lnZ obtained in the simulation in which
the
construction of the C-matrix was started when $lnf$ becomes just less than 
$10^{-2}$
. Following this the simulation was continued till $lnf$ reaches $10^{-9}$
, with the usual procedure of DOS refreshing along with the condition of 90\% 
histogram flatness. It is clear from the figure that in order to get the
best result, a compromise is to be made in getting the most useful C-matrix.
If it is constructed from the very beginning of the simulation while the DOS
profile is far away from being a realistic representation of the actual one,
 the errors which accumulate in the C-data lead to large error. Again
when the construction begins at a stage too late, some important information
of the simulation process is lost resulting again in increase in errors. In 
particular, figure 2 shows that the error for $\lambda_2$ is significantly 
smaller 
 than that for $\lambda_0$.
Also shown in figure 2 is the error in $lnZ$ for pure WL simulation
with the condition of $90\%$ histogram flatness imposed. It is clear from
the diagram that it is possible to achieve better accuracy with the WLTM 
 algorithm. However the gain in the accuracy seems to be only marginal and 
that too is at an appreciable increase in the CPU time involved.\\
%%%%%%%%%%%%%%%%%%%%%%%%%%%%%%%%%%%%%%%%%%%%%%%%%%%%%%%%
%%%%%%%%%%%%%%%%%%%%%%%%%%%%%%%%%%%%%%%%%%%%%%%%%%%%%%
\begin{figure}[htbp]
\begin{center}
\resizebox{120mm}{!}{\rotatebox{0}{\includegraphics[scale=.45]{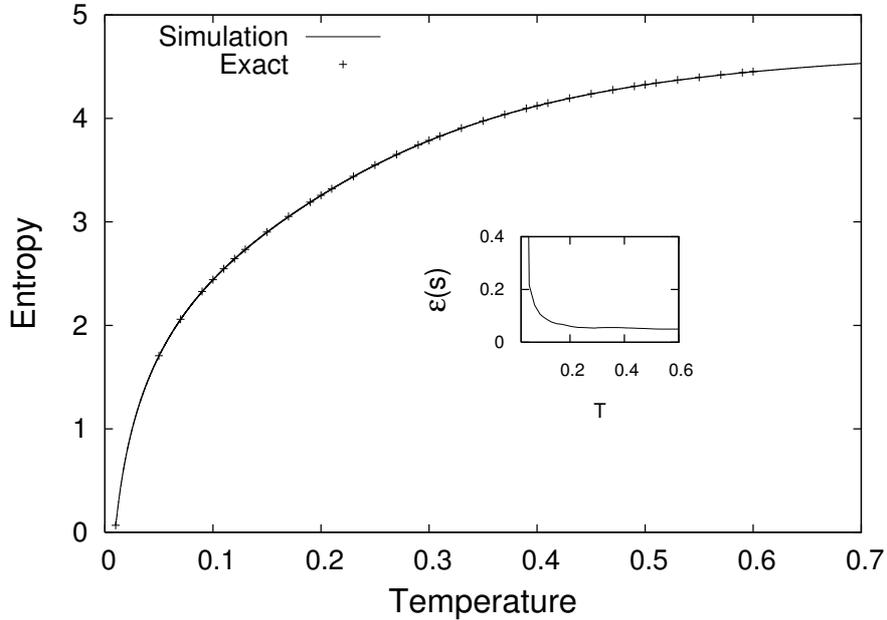}}}
\end{center}
\caption{ The variation of entropy per particle
with temperature for $L=220$ 1-d LL model for $\lambda=\lambda_0$ is shown.
Exact results are
also plotted and the error in entropy obtained in comparison with the exact
result is shown in the inset.}
\label{fig7.eps}
\end{figure}
%%%%%%%%%%%%%%%%%%%%%%%%%%%%%%%%%%%%%%%%%%%%%%%%%%%%%%%%
\noindent In figure 3 is shown the error in $lnZ$ for the three lattice 
sizes, achieved with the WLTM algorithm and C-matrix updating corresponding
to $\lambda_{2}$. Since the theoretical estimate 
of $lnZ$ \cite{romerio} is for the
thermodynamic limit, it is natural to expect that the error should decrease with
the increase in system size.
In figure 4, we have compared the WLTM results
obtained with the $90\%$ flatness method with those obtained with the 'single
visit per bin' condition imposed. It is clear that the latter leads to a
huge error in the density of states and hence in the partition function and
is not usable in a continuous model.
In figure 5, 6 and 7 we have plotted the
 energy, specific heat and entropy against temperature for L=220 for 
$\lambda_0$-simulation and the results have been compared with the exact
 values of these observables obtained from the exact results 
\cite{romerio}. In the
inset in each figure we have also shown the corresponding errors in these
parameters obtained in comparison with the exact results.\\
\subsection {\bf The 2-d XY model}

\noindent In this model simulations were carried out for lattice sizes
$10 \times 10$ and $20 \times 20$ and the corresponding minimum of the
energy range were chosen to be 3 for each. Histogram 
flatness was restricted
to $80\%$ in all WL and WLTM simulations. In addition to the already mentioned
cut in the energy range, for the $20\times 20$ system, in order to save CPU
time, we had to relax the condition for histogram flatness 
check for the first 500 bins.
The XY-model and the phase
transition it exhibits have been worked out in detail in references
\cite{KT, K}. However it is not possible to compare
the results of our simulation in the XY-model for the DOS or partition function,
with the exact results for this model, as we have done for the 1-d LL model,
because such exact results for this model are not available.
%%%%%%%%%%%%%%%%%%%%%%%%%%%%%%%%%%%%%%%%%%%%%%%%%%%%%%
\begin{figure}[htbp]
\begin{center}
\subfigure[]{\resizebox{120mm}{!}{\rotatebox{0}{\includegraphics[scale=.5]{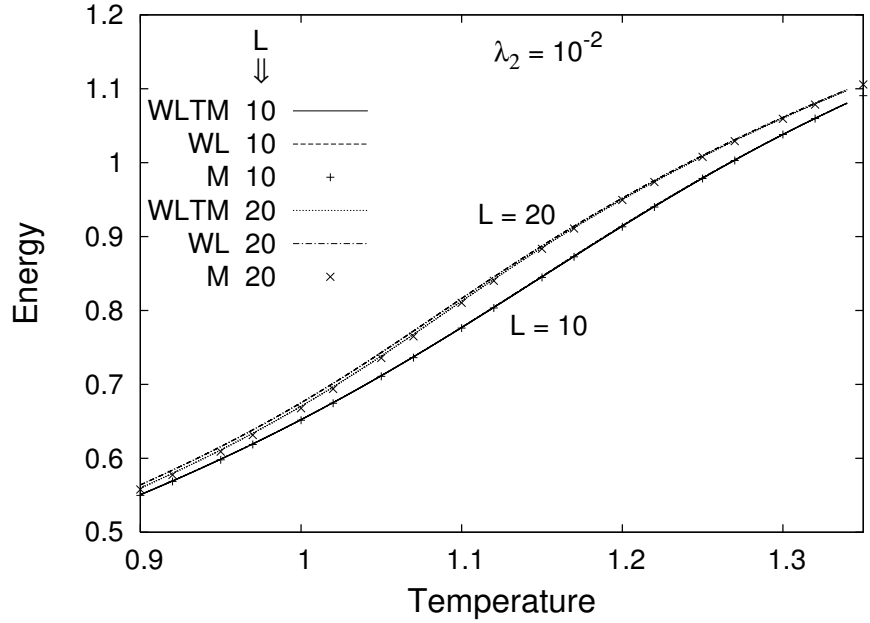}}}}
\subfigure[]{\resizebox{120mm}{!}{\rotatebox{0}{\includegraphics[scale=.5]{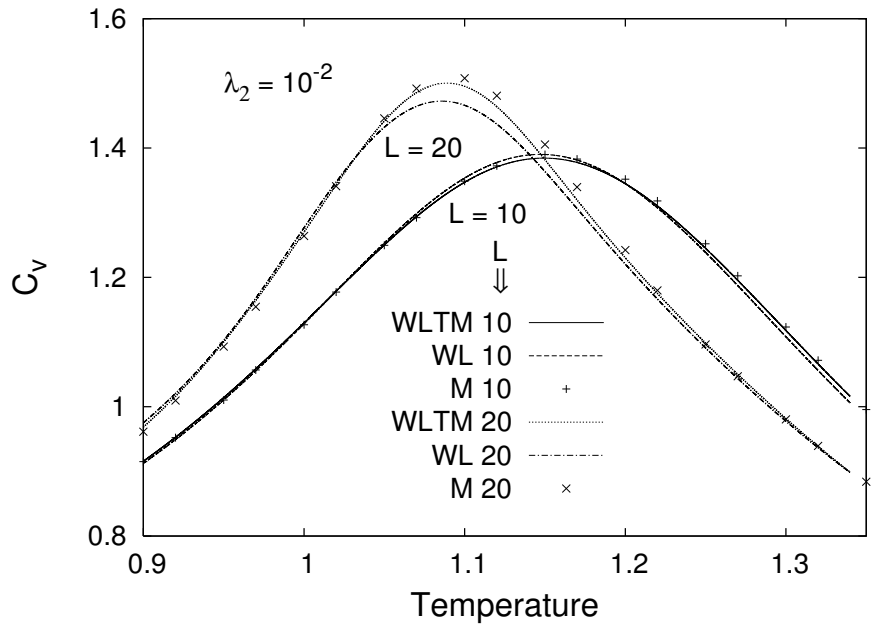}}}}
\end{center}
\label{fig8a.eps}
\caption{ The average energy per particle for the XY-model
is plotted (a) against
temperature for $10\times 10$ and $20\times 20$ lattice sizes, obtained
using three MC methods namely, WLTM, WL and M (Metropolis). These
results correspond to the $\lambda_2$ case and all results are obtained
by taking averages over 20 independent simulations.
Similar plots for the specific heat per particle against temperature are shown
in (b).
}

\end{figure}
%%%%%%%%%%%%%%%%%%%%%%%%%%%%%%%%%%%%%%%%%%%%%%%%%%%%%%%%
We therefore present here a 
comparison of our WL and WLTM results with those obtained from simulation
using the conventional Metropolis algorithm, which is known to work 
satisfactorily for this model \cite{palma}. To increase the 
reliability of our results,
we have performed 20 independent simulations for each of 
the Metropolis, WL and WLTM method and have averaged the 
respective results. In the Metropolis simulation the averaging was done 
directly over the observables while in the other two cases, the DOS was
averaged before it was used to obtain the observables.\\ 
%%%%%%%%%%%%%%%%%%%%%%%%%%%%%%%%%%%%%%%%%%%%%%%%%%%%%%%%
\noindent Regarding the choice of energy bin-width $d_e$ and the parameter
$p$, the considerations as were applied to the 1-d LL model, were
taken into account. Unless otherwise stated, we have used $d_e=0.1$ and 
$p=0.1$ in this model too. This led us to work with a large number of bins,
namely 8000, for the $20 \times 20$ lattice.\\
\noindent In figures 8(a) and 8(b) we have shown the variation
of the average energy, $E$ and the specific heat, $C_v$ respectively with 
the temperature
%%%%%%%%%%%%%%%%%%%%%%%%%%%%%%%%%%%%%%%%%%%%%%%%%%%%%%
\begin{figure}[htbp]
\begin{center}
\subfigure[]{\resizebox{120mm}{!}{\rotatebox{0}{\includegraphics[scale=.5]{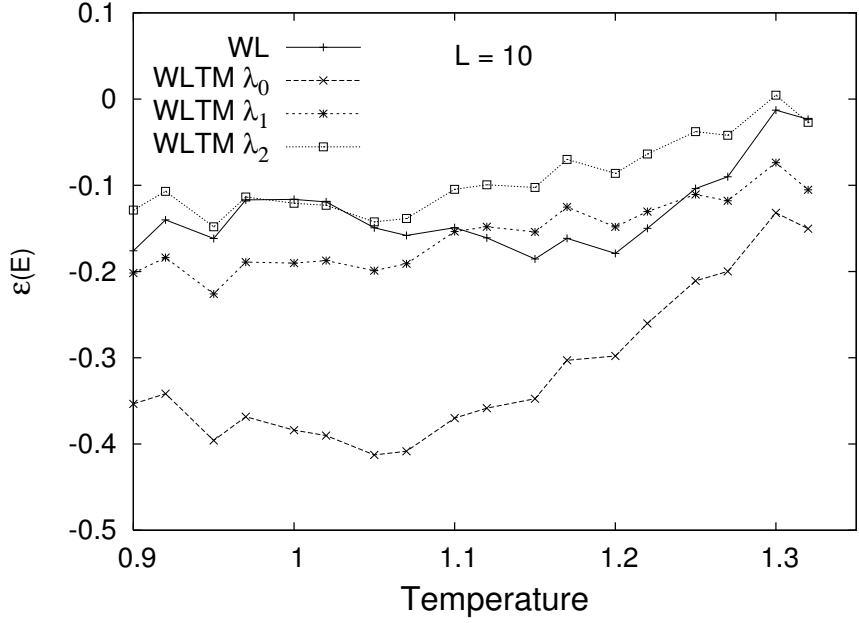}}}}
\subfigure[]{\resizebox{120mm}{!}{\rotatebox{0}{\includegraphics[scale=.5]{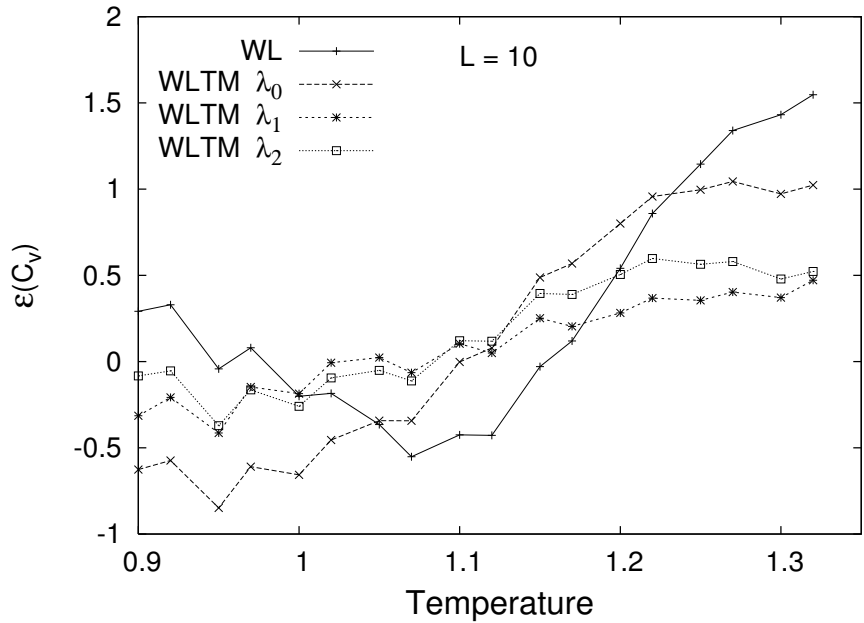}}}}
\end{center}
\caption{ This figure (a) shows the errors in energy for the
XY-model for $10\times 10$ lattice obtained from WL simulation and
WLTM simulation for three different starting points
$\lambda_0, \lambda_1, \lambda_2$ of the C-matrix. The errors have
been calculated by comparing the results of simulations with that obtained
by using the Metropolis algorithm. The straight line segments are guide
to the eyes.
Similar plots for the
specific heat per particle against temperature are shown in figure (b).}

\label{fig9}
\end{figure}
%%%%%%%%%%%%%%%%%%%%%%%%%%%%%%%%%%%%%%%%%%%%%%%%%%%%%%%%
%%%%%%%%%%%%%%%%%%%%%%%%%%%%%%%%%%%%%%%%%%%%%%%%%%%%%%
\begin{figure}[htbp]
\begin{center}
\subfigure[]{\resizebox{120mm}{!}{\rotatebox{0}{\includegraphics[scale=.5]{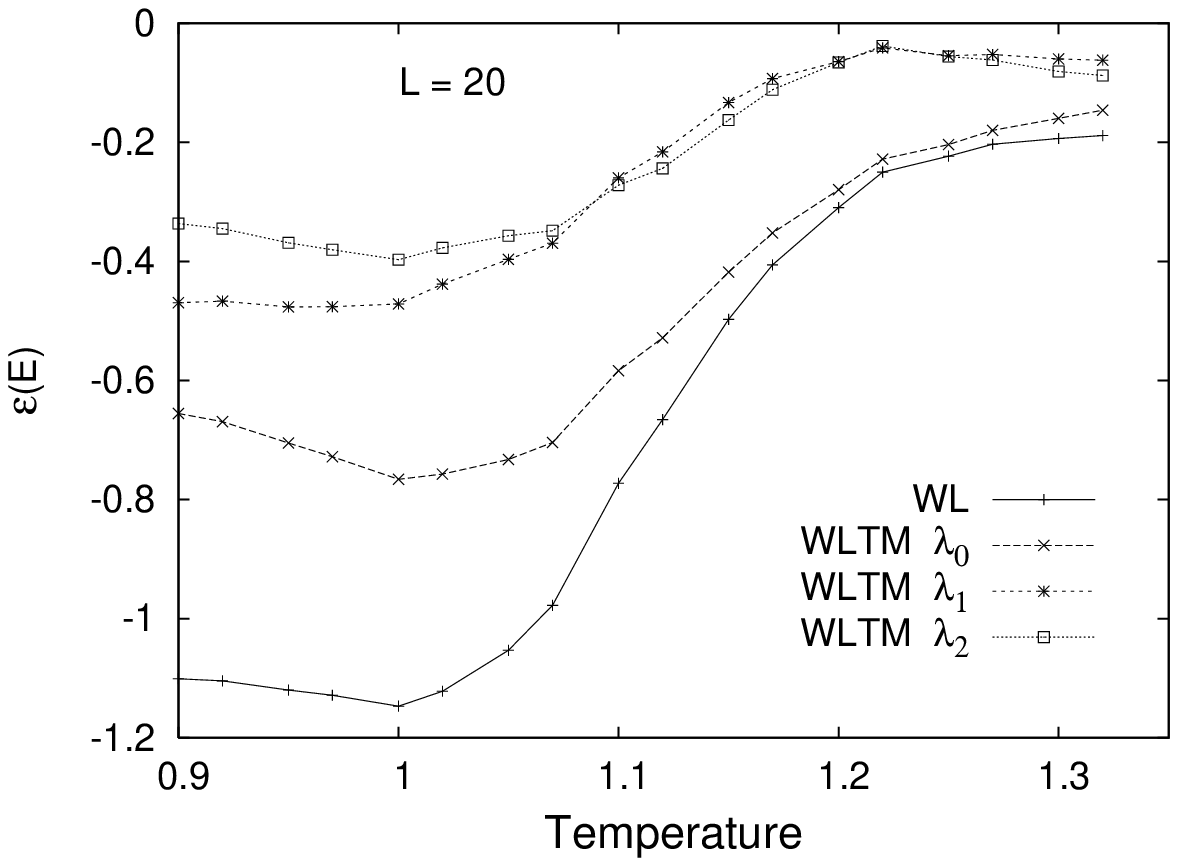}}}}
\subfigure[]{\resizebox{120mm}{!}{\rotatebox{0}{\includegraphics[scale=.5]{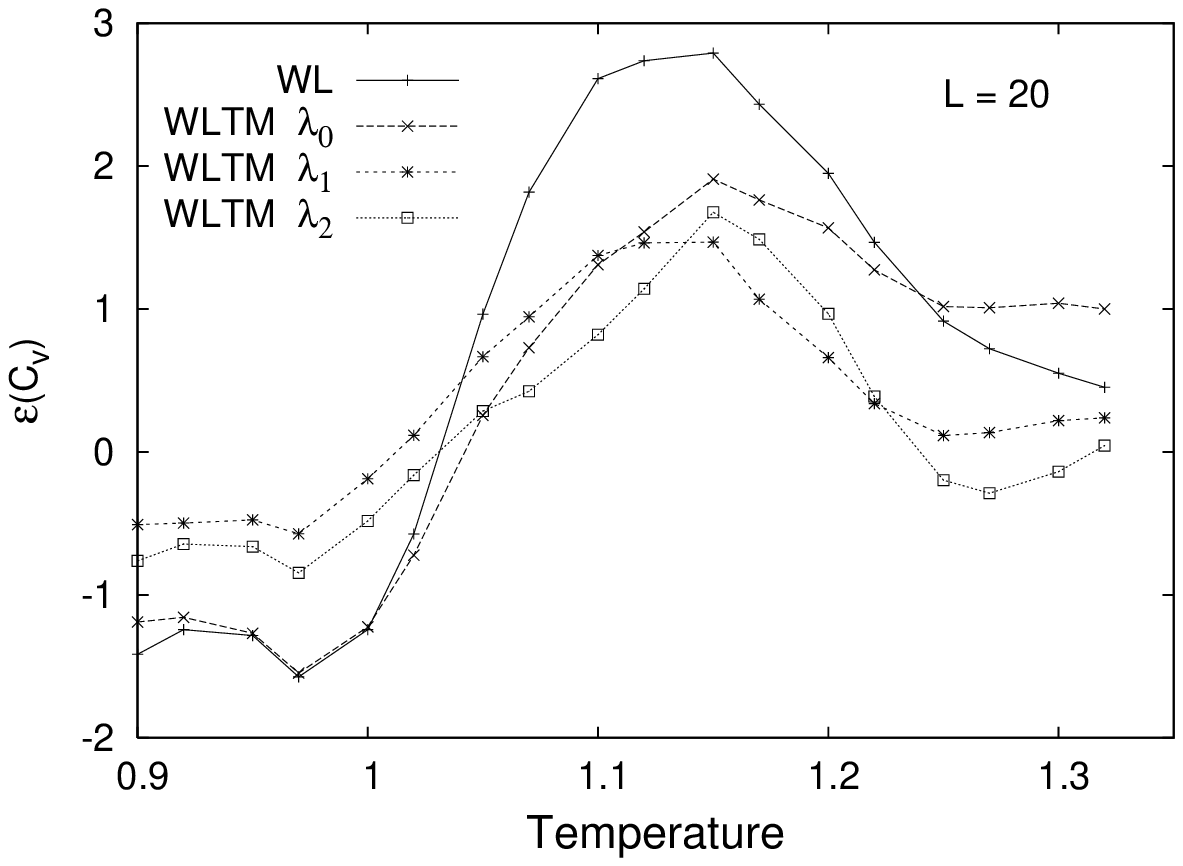}}}}
\end{center}
\caption{ Plots are same as those in Figure \ref{fig9} for the
$20\times 20$ lattice sizes.}

\label{fig10a.eps}
\end{figure}
%%%%%%%%%%%%%%%%%%%%%%%%%%%%%%%%%%%%%%%%%%%%%%%%%%%%%%%%
for the $10 \times 10$ and $20 \times 20$ lattices, obtained with the WL,
WLTM and Metropolis algorithms. This diagram and all the following diagrams,
depicting the results of the XY-model, represent the results of 20 independent
simulations for each of the algorithms used. As the plots shown in figure
8 do not clearly reveal the errors in the different simulations, we 
present these in detail in figure 9. The percentage errors presented in figures
9 to 12 have been obtained by comparing the results of simulations with that
obtained by using the Metropolis algorithm. Figure 9(a) shows the errors
in energy for the $10\times 10$ lattice obtained for the pure WL simulation
and the WLTM simulation for three different starting points of the C-matrix
. Figure 9(b) shows the same for the specific heat $C_v$. For the energy
curve, WLTM simulation with $\lambda_{2}$ seems to be the best choice where
the error is $\sim 0.1\%$. In the specific heat curve, where the errors are
slightly greater, as is to be expected for this being a fluctuations
quantity, simulations with $\lambda_{1}$ and $\lambda_{2}$ 
have comparable errors.
It is clear from these diagrams that the choice of $\lambda_{0}$ for the
commencement of the C-matrix leads to larger errors in the WLTM simulation.
Figures 10(a) and 10(b) are the plots of the errors in energy and specific
heat for the $20\times 20$ lattice. These diagrams clearly reveal that the
errors in the pure WL algorithm are greater than those for the WLTM algorithm
and for the latter, the performance of the $\lambda_{0}$ simulation is worst.
%%%%%%%%%%%%%%%%%%%%%%%%%%%%%%%%%%%%%%%%%%%%%%%%%%%%%%
\begin{figure}[htbp]
\begin{center}
\resizebox{120mm}{!}{\rotatebox{0}{\includegraphics[scale=.5]{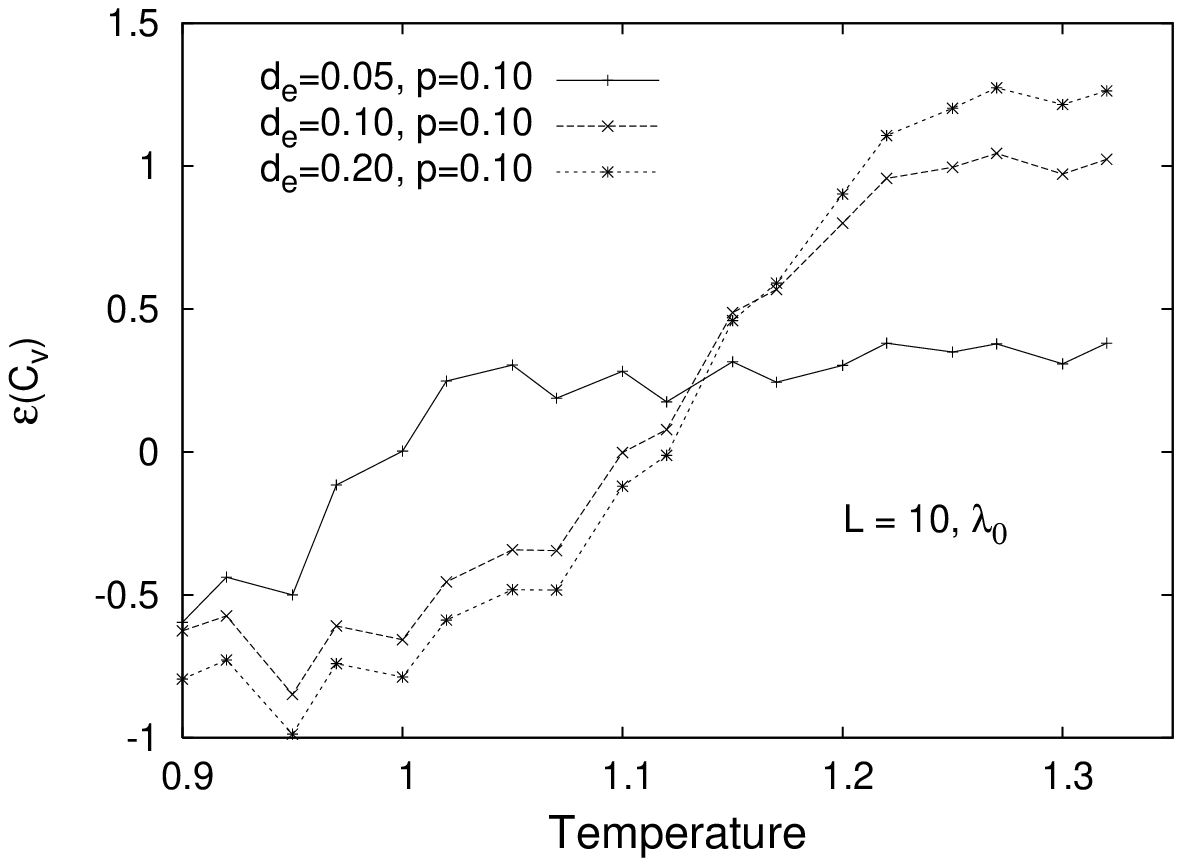}}}
\end{center}
\caption{ The errors in specific heat $C_v$, for the different
choices of the bin width keeping other parameters fixed, are plotted
against temperature.}
\label{fig11.eps}
\end{figure}
%%%%%%%%%%%%%%%%%%%%%%%%%%%%%%%%%%%%%%%%%%%%%%%%%%%%%%%%
In figure 11 we have shown the errors in specific heat $C_v$ for different
choices of the bin width $d_{e}$, other conditions remaining the same.
As is perhaps to be expected, the smaller bin-width $d_{e}=0.05$ gives
best results while the choice $d_{e}=0.2$ is the worst. These results
are for the $\lambda_{0}$ simulation. Figure 12 demonstrates that the
$\lambda_{0}$ simulation with $d_{e}=0.05$ and the $\lambda_{1}$ simulation
for $d_{e}=0.1$ leads to comparable errors in $C_v$.
%%%%%%%%%%%%%%%%%%%%%%%%%%%%%%%%%%%%%%%%%%%%%%%%%%%%%%
\begin{figure}[htbp]
\begin{center}
\resizebox{120mm}{!}{\rotatebox{0}{\includegraphics[scale=.5]{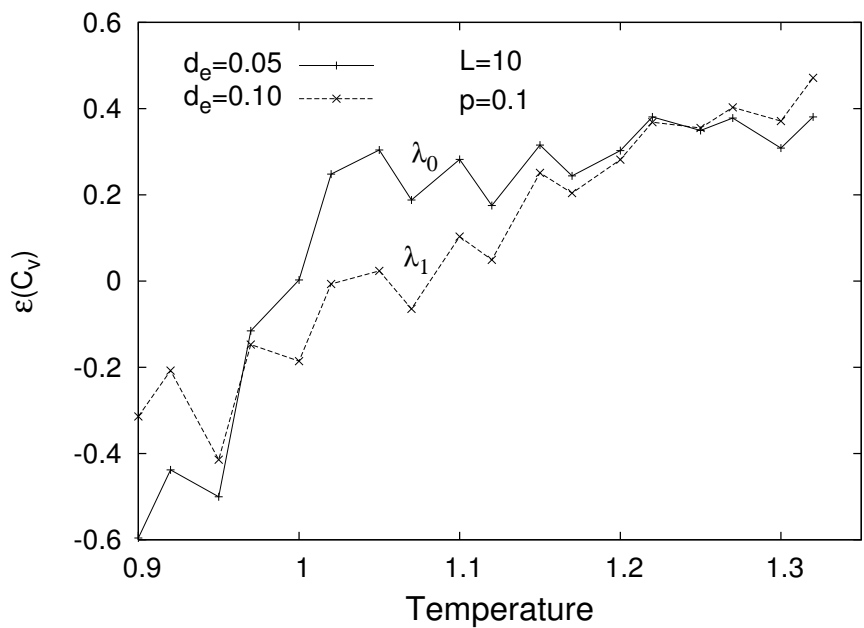}}}
\end{center}
\caption{ A comparison of the errors of $\lambda_0$
simulation with bin width $d_e=0.05$ and the $\lambda_1$ simulation with
bin width $d_e=0.1$.}
\label{fig12.eps}
\end{figure}
%%%%%%%%%%%%%%%%%%%%%%%%%%%%%%%%%%%%%%%%%%%%%%%%%%%%%%%%
\section{\bf Conclusion}
\noindent We have presented results of extensive MC simulations 
using WL and WLTM
algorithms in two continuous lattice spin models. Simulation of 
continuous models needs a choice of two parameters, a bin-width $d_e$
(for energy) and $p$, which determines the amplitude of the rotation to be 
imparted to the spins. We are of the opinion that while the bin-width
should be as small as possible, depending on the computer resources
available, the parameter determining the amplitude of spin rotation
should be chosen in such a way that, of the proposed moves during the
random walk, about $50\%$ should take the random walker out of the
starting macrostate to a neighbouring one. In the 
simulation of a model involving
continuous macroscopic variables, these two parameters critically 
determine the accuracy of the results, while no such 
considerations are needed for simulating a discrete model.\\
\noindent Our work on the two continuous models show that it is possible
to obtain results of better accuracy with the WLTM algorithm than that is
obtainable from the WL algorithm. For this purpose, one must be careful
about the starting point for the construction of the C-matrix. While this 
matrix should contain as much information as possible regarding the 
transitions proposed, care must be taken not to include information 
about the proposed moves which are generated very early in the simulation,
when the profile of the DOS is just in a formative stage. We had also
to make an improvisation in the WLTM algorithm regarding the question of
convergence of the DOS. The 'single visit to each macrostate' criterion,
which saves
a lot of CPU time and was proposed by Shell {\it et. al} in their original
work \cite{shell}, though adequate for a discrete model, seems to result in poor
accuracy of the results for a continuous model. So we had to retain
the condition for the flatness check of the histogram as is used in the 
original version of the WL algorithm. This of course results in a significant
increase in the CPU time.\\
\noindent We end this section with a few comments. We are of the opinion
that the gain in accuracy over the WL algorithm which can be achieved using
the WLTM algorithm, with a choice of various parameters as well as the 
commencement point of the C-matrix, done in the most judicious way, is not
significant. This too is possible only at the cost of a huge increase 
in CPU time. We would also point out that simulating a continuous model
of a reasonably large size using the WLTM algorithm, leads to the 
requirement of a vast amount of computer memory (RAM) and it would be an 
almost
impossible task if one needs to determine a joint density of state by 
performing a two-dimensional random walk ( including quantities like
order parameter, correlation function etc. besides energy ) in a
system of a reasonably large size. It may also be added that for this 
purpose, the WL algorithm has also 
proved to be inadequate \cite{kisor, suman}.\\
\noindent We are of the view that in order to simulate a continuous model
of large size, the more than a decade old method of multiple
histogram reweighting, proposed by Ferrenberg and Swendsen 
\cite{ferren1,ferren2} still 
seems to be the best choice. In addition, if one is dealing with a system
not exhibiting a temperature induced first order phase transition, the
Wolff cluster algorithm \cite{wolff} is useful for reducing critical 
slowing down. Examples of such work, to name a few, may be found in
\cite{enakshi, mouritsen, kunz, abhijit1, abhijit2}.
The methods like WL, TM or WLTM which directly determine the DOS 
do not seem to be a suitable choice for these jobs, unless of course one uses
a computer code for parallel processing using OPENMP or MPI \cite{zhan}.
\section{Acknowledgment}
\vskip .1in
\noindent We acknowledge the receipt of a research grant No.
03(1071)/06/EMR-II from Council of Scientific and Industrial Research (CSIR),
India which helped us to procure the IBM x226 servers. One of the authors (SB)
gratefully acknowledges CSIR, India, for the financial support.

%%%%%%%%%%%%%%%%%%%%%%%%%%%%%%%%%%%%%%%%%%%%%%%%%%%%%%
%\begin{figure}[tbh]
%\begin{center}
%\resizebox{100mm}{!}{\rotatebox{0}{\includegraphics[scale=.3]{fig2.eps}}}
%\end{center}
%\caption{The entropy per particle with temperature for lattice sizes
%$L=220$ is plotted. The exact result is also shown in the figure. The
%relative errors are shown in the inset.  }
%\label{entpy1.eps}
%\end{figure}
%%%%%%%%%%%%%%%%%%%%%%%%%%%%%%%%%%%%%%%%%%%%%%%%%%%%%%%%

\end{document}